\documentclass[aps,pra,showpacs,twoside,twocolumn,10pt]{revtex4-1}
\usepackage[colorlinks=true, citecolor=red, urlcolor=blue ]{hyperref}
\usepackage{epsfig,newlfont,amssymb,amsfonts,amsmath,bm,subfigure,palatino,mathtools,amsthm,braket,times,soul}
\usepackage{multirow}

\begin{document}
\begin{center}
\title{Restrictions on shareability of classical correlations for random multipartite quantum states}

\author{Saptarshi Roy}
\affiliation{Harish-Chandra Research Institute and HBNI, Chhatnag Road, Jhunsi, Allahabad - 211019, India}

\author{Shiladitya Mal}
\affiliation{Harish-Chandra Research Institute and HBNI, Chhatnag Road, Jhunsi, Allahabad - 211019, India}

\author{Aditi Sen(De)}
\affiliation{Harish-Chandra Research Institute and HBNI, Chhatnag Road, Jhunsi, Allahabad - 211019, India}

\begin{abstract}

Unlike quantum correlations, the shareability of classical correlations (CCs) between two-parties of a multipartite state is assumed to be free since there exist states for which CCs for each of the reduced states can simultaneously reach their algebraic maximum value. However, when one randomly picks out states from the state space, we find that the probability of obtaining those states possessing the algebraic maximum value is vanishingly small.  Therefore, the possibility of a nontrivial upper bound on the distribution of CCs that is less than the algebraic maxima emerges. We explore this possibility by Haar uniformly generating random multipartite states and computing the frequency distribution for various CC measures,  conventional classical correlators,  and two axiomatic measures of classical correlations, namely the classical part of quantum discord and local work of work-deficit. We find that the distributions are typically Gaussian-like and their standard deviations decrease with the increase in number of parties.    It also reveals that among the multiqubit random states, most of the reduced density matrices possess a low amount of CCs which can also be confirmed by the mean of the distributions, thereby showing a kind of restrictions on the shareability of classical correlations for random states. Furthermore, we also notice that the maximal value for random states is much lower than the algebraic maxima obtained for a set of states, and the gap between the two increases further for states with a higher number of parties.  We report that for a higher number of parties, the classical part of quantum discord and local work can follow monogamy-based upper bound on shareability while classical correlators have a different upper bound. The trends of shareability for classical correlation measures in random states clearly demarcate between the axiomatic definition of classical correlations and the conventional ones.
\end{abstract}
\maketitle
\end{center}

\section{Introduction}
In a multipartite system,  the rule according to which certain physical property is shared among different subsystems is  assigned by a specific theory. 
In particular, for a given theory which can be quantum mechanics \cite{qmbook} or generalized probabilistic theory \cite{prbox}, the physical characteristics, say, \(\mathcal{P}\), of reduced states  of a multipartite state, \(\rho_{1 \ldots N}\) shared by \(N\) parties  situated at different locations  can be  upper bounded by a fixed value, thereby establishing the restrictions on shareability of that physical component. The mathematical formulation of it reads as 
\begin{eqnarray}
\label{eq_sharingcond}
\sum_{i = 2}^N \mathcal{P} (\rho_{1i}) \leq U,
\end{eqnarray}
where \(\rho_{1i}\) is the reduced state of  \(\rho_{1 \ldots N}\) and \(U\) is an upper bound of the shareability condition. Like  no-go theorems for single quantum systems \cite{nocloning, nobroadcast, nodeleting, nobit, nogoetc, nomask},  constraints  proved on  sharing of  properties like entanglement, violation of Bell inequalities, capacities of dense coding and teleportation  in a multipartitie quantum system \cite{CKW, sg'2001, bhk'2005, tv'2006, mag'2006, telemono, t'2009, pb'2009, ow'2010, kprlk'2011, agca'2012, foundation-mono3, exclu, corrnet} play an important role in quantum information processing tasks.

The  unbounded sharing of quantum correlations among  a pair of parties in a multipartite state is forbidden --  a concept known as monogamy of quantum correlations (QC) \cite{CKW, review}. In particular,  if two of the parties of a multipartite state share maximal  QC, they cannot share any QC with other parties.  Monogamy of QC also has an impact on several quantum information processing tasks which include quantum cryptography, entanglement sharing in a quantum network \cite{terhal, cryptorev, teleportation}.  
In a seminal paper by Coffman-Kundu-Wootters \cite{CKW}, such a  qualitative concept of monogamy got a mathematical form that can be used to check whether a QC measure follows a monogamy inequality or not. Specifically,  a QC measure, \(\mathcal{Q}\), is said to follow a monogamy relation \cite{monogamyscore, ent_monogamy}, if 
\begin{eqnarray}
\label{eq_monogamyQC}
\delta_{\mathcal{Q}} = \mathcal{Q}_{1:2\ldots N} - \sum_{i=2}^{N} \mathcal{Q}_{1i} \geq 0,
\end{eqnarray}
where \(\mathcal{Q}_{1:2\ldots N} \equiv \mathcal{Q}(\rho_{1:2 \ldots N})\),  \( \mathcal{Q}_{1i} \equiv \mathcal{Q}(\rho_{1i}), \, i =2, \ldots N\), of a multipartite state \(\rho_{12\ldots N}\) and \(\delta_{\mathcal{Q}}\)  can be referred as QC monogamy score \cite{monogamyscore}.  In other words, although each term in \(\sum_{i=2}^{N} \mathcal{Q}_{1i}\) can reach \(\log_2 d\) (excepting measures like negativity and logarithmic negativity \cite{neg}) in a \(N\)-qudit system, sharing of QC is bounded above only by  a quantum correlation  content in the \(1:\mbox{rest}\)-bipartition.  Note that  the party, \(1\), has a special status and can be referred to as a nodal observer. Similar to such inequality can also be derived with other party as a nodal observer. It is known that monogamy scores of squared concurrence \cite{concurrence, CKW}, negativity \cite{neg, logneg, negativitymonogamysq}, quantum discord \cite{ discord1, discord2, discordrev, foundation-mono3, dis_sq, zurek}  are nonnegative. Moreover, it was shown that  all QC measures for random multipartite quantum states tend to become monogamous when the number of parties increases \cite{Eisertrand, Winterrand, Sooryarand, Ratulrand}.

In stark contrast, classical correlations (CCs) do not possess such restrictions. Specifically, there exists a multipartite state for which any CC quantifier between reduced two party states  can simultaneously reach its maximal value and hence the upper bound in Eq. (\ref{eq_sharingcond}) scales with the increase in the number of parties.  However, it should be noted that unlike QC measure, it is not yet settled when a quantity can measure reliably the amount of CC present even in a bipartite quantum state. Over the years,  a few measures of CC  were proposed  -- prominent ones  having  diverse origins include CC part in  quantum discord (CQD) \cite{ discord1, discord2, discordrev}, extractable local work (LW) \cite{wd}, and a conventional classical correlators (CCC),  defined as \(\mbox{tr} (\sigma^{k} \otimes \sigma^{l} \rho_{12})\) for a bipartite state, \(\rho_{12}\) which have been used in quantum mechanics, ranging from Bell inequalities \cite{HorodeckiBellin, Bellreview} to many-body physics \cite{Sachdev}. CQD and LW are defined operationally and satisfy some axioms which a bona fide measure of CC is supposed to obey.

In this paper, we address the following questions --\\
 \emph{Can we obtain a non-trivial upper bound ( \(U\) in Eq. (\ref{eq_sharingcond})) on the shareability of CC  among bipartite reduced states of random multipartite systems?  \\
  Secondly, how does the frequency distribution, and consequently the bound for  sharing of CC among bipartite reduced states obtained from random multipartite states change with the increase in system-size?} \\
  We report here that the answer to the first question   is affirmative, and hence a new rule for the shareability of CC among subsystems emerges for random multipartite states.  Investigating on Haar uniformly generated random multipartite states \cite{Karolbook}, we find several counter-intuitive results. 
For  systematic analysis, the shareability for classical correlations is addressed  from two perspectives which we refer to as   ``unconstrained" and ``constrained'' settings. The ``constrained" one implies that the sample of random states that we choose for our analysis possesses a fixed, or a definite range of values of a particular physical property (classical or quantum) different from the one under investigations while the unconstrained one does not have such restrictions. By carrying out our investigations 
for $N = 3$ to $6$ multi-qubit random pure states, we observe that like QC, maximal shareability of CC is also restricted, rather the algebraic maximum occurs only for  sets of states with vanishingly small measure. In the case of an unconstrained scenario, the frequency distributions of the shareability constraints for random states (i.e., the left hand side in Eq. (\ref{eq_sharingcond})) take the form of a Gaussian, irrespective of the choices of the CC measures and the Gaussian-like shapes become  narrower for higher number of qubits, thereby showing the decrease in standard deviation with the increase of number of parties. On the other hand,  the mean value of CCC remains almost constant over increasing system-size, while the means of CQD and LW decrease.  Moreover, their maximum  values obtained via numerical simulations decrease with the increase in the number of parties. We also find a kind of trade-off for  maximal values of CCCs in complementary directions.

In the case of a constrained framework,  we consider two kinds of constraints -- for a definite value of  CCC in a fixed direction, we study the behavior of sharing rule for CCC in complementary direction and  we also investigate the consequence on  average as well as the maximum value  of \(\sum_{i = 2}^N \mathcal{P}_{1i}\) for the  CC measures when randomly generated states possess a definite range of  genuine multipartite entanglement. Interestingly, we notice that with the increase of genuine multipartite entanglement, the average value for the shareability of CQD and LW in the subsystems of random multipartite states diminishes. Such an observation leads to the result that
LW and CQD follow the monogamy-based  upper bound with a very high percentage of random states having a higher number of parties which CCC fails to satisfy.

The paper is organized in the following way. In Sec. \ref{sec_CCmeasure}, we discuss the classical correlation measures, and the class of states for which shareability of CC measures reach their maximum value. Sec. \ref{sec:uncon1} deals with the patterns in the distribution of CC in multiqubit random states while we discuss how the sharing properties of CC changes when a fixed amount of other CC measure or a genuine multipartite entanglement measure  is present in random states in Sec. \ref{sec:con1}.  We check whether the monogamy-motivated upper bound on shareability of CC measures is good or not in  Sec. \ref{sec_boundingbymonogamy}, and   conclude in Sec. \ref{sec_conclu}.

\section{Classical correlation measures and their algebraic maxima}
 \label{sec_CCmeasure}
Let us describe briefly three types of classical correlation  (CC) measures and their properties for an arbitrary bipartite shared state, \(\rho_{12}\).  
Unlike entanglement measures \cite{horodecki}, the properties that a ``good" classical correlation measure of quantum states  should follow are not well understood. However, there are CC measures introduced   in \cite{discord1, discord2, discordrev} which follow the following properties -- (1)  it should be vanishing for \(\rho_1 \otimes \rho_2\); (2) it is invariant under local unitary transformations; (3) it should be non-increasing under local operations and (4) it reduces to \(S(\rho_1) = S(\rho_2)\) for pure bipartite states, \(|\psi\rangle_{12}\), with \(\rho_1\) and \(\rho_2\) being the corresponding local density matrices.
We will also consider another CC measure  introduced from the perspective of thermodynamics and the conventional classical correlators, apeeared in the definition of density matrices \cite{HorodeckiBellin},  which play an important role in dfifferent fields ranging from  Bell inequalities \cite{Bellreview} to many-body physics \cite{Sachdev} (see also \cite{qwtcl, HoroBennett}).


%
 
   


We first give the definitions of two  classical correlation measures  \cite{discordrev}  associated with quantum discord (QD) and  one-way work-deficit where the former do follow the postulates of CC measure  while the latter satisfies the  first two  and the third one with modifications. We refer both these CC measures as the axiomatic ones. The  classical correlation part of quantum discord  (CQD) of \(\rho_{12}\)  can be defined   as 
\begin{eqnarray}
\label{eq_CCQD}
C^D (\rho_{12}) \equiv C^D_{12} = S(\rho_1)  - \min_{\{P_i\}} \sum_i p_i  S(\rho_{1|i}),  
\end{eqnarray}
where  \( S(\rho) = - \mbox{tr} (\rho \log_2 \rho) \) is the von Neumann entropy,    
\begin{equation}
\label{eq_condirh}
\rho_{1|i} = \frac{\mbox{tr}_2 (I \otimes P_i \rho_{12} I \otimes P_{i})}{\mbox{tr} (I \otimes P_i \rho_{12} I \otimes P_{i})}, 
\end{equation}
 with \(P_i\) being the rank-1 projective measurements on the second  party and \(p_i= \mbox{tr} (I \otimes P_i \rho_{12} I \otimes P_{i})\). Here the minimization is performed over all rank-1 projective measurements. Similar definition emerges when measurement is done on the first party. Notice that  in the definition of QD, the optimization is taken over the most general measurements, i.e, positive operator valued measurements (POVMs). However, it was shown via numerical simulations that projective measurements yield very close to the optimal value obtained via POVMs. So from the practical viewpoint of computational simplicity, we perform our analysis with projective measurements.  
 
Motivated by quantum thermodynamics, the classical correlation can also be quantified as  local extractable work (LW) by closed local operations and one-way classical communication  \cite{discordrev, wd} consisting of local unitaries, local dephasings, and sending dephased states from one party to another. Mathematically, LW reads as
\begin{eqnarray}
\tilde{C}^{LW} (\rho_{12}) \equiv  \tilde{C}^{LW}_{12} = \log_2 d_{12} - \min_{\{P_i\}} S(\sum_i p_i \rho_{1|i}),
\label{eq_lwdef}
\end{eqnarray} 
where \(p_i\) and \( \rho_{1|i}\) are same as in Eq. (\ref{eq_condirh}), and \(d_{12}= d_1 d_2\) is the dimension of \(\rho_{12}\) with the individual subsystems having dimensions, \(d_1\) and \(d_2\). 
Note that $\tilde{C}^{LW}$ can take values upto $\log_2 d_{12}$ and to make it consistent with other measures of classical correlation, which take values from $0$ to $1$, we scale $\tilde{C}^{LW}$ with $\log_2 d_{12}$ and call it as 
$C^{LW} = \frac{1}{\log_2 d_{12}}\tilde{C}^{LW}$.


Let us now define  conventional  two-site  classical correlator present in any two-qubit state, given by 
\begin{eqnarray}
 \rho_{12}  &=&  \frac{1}{4} (I\otimes  I + \sum_{k = x, y, z} ( m^{k} \sigma^{k} \otimes I   + m'^{k} I \otimes \sigma^k \nonumber \\ 
 &+& \sum_{k, l = x, y, z}  C^{k l} \,  \sigma^k \otimes \sigma^{l}). 
\end{eqnarray} 
Here 
\begin{eqnarray}
C^{k l} = \mbox{tr} (\sigma^{k} \otimes \sigma^{l} \rho_{12}), \, k, l = x, y, z.
\label{eq_CCdef}
\end{eqnarray} 
represents the two-site classical correlators which leads to the correlation matrix having diagonal elements \(C^{kk}, k =x, y, z\) and off-diagonal ones, \(C^{k l}, k \neq l\).  \(m^{k} = \mbox{tr} (\sigma^{k} \rho_{1})\), \(m'^k = \mbox{tr} (\sigma^{k} \rho_{2}), \, k =x, y, z\)s denote the magnetizations corresponding to  the single site density matrix of \(\rho_{12}\). Note that  \(C^{k l}\) does not follow the  properties mentioned above and hence we may expect to see different universal behavior for random states than that of \(C^D\) and \( \tilde{C}^{LW}\).  Since the classical correlators varies from \(-1\) to \(1\), we scale its range from \(0\) to \(1\), by taking the absolute value of the same. Since from now on, we will always use the absolute values of these correlators, we drop the absolute bars, and any reference to $C_{1i}^{kl}$ means the absolute value of the quantity, unless mentioned otherwise.


As stated earlier, we aim to investigate the pattern in the distributions of  \(\sum_{i=2}^N C^D_{1i}\),  \(\sum_{i=2}^N C^{LW}_{1i}\), and  \(\sum_{i=2}^{N} C_{1i}^{kl},\) as well as their non-trivial upper bounds  for random  multipartite states,  \(\rho_{12 \ldots N}\)  by varying the number of parties. We are also interested to compute the corresponding statistical quantities like different moments of the distributions and compare them. 
 Unlike QCs, we first notice that each quantity in the sum can simultaneously take the  maximum value, unity for qubits.  In the next subsection, we will identify  classes of multipartite states for which the algebraic maxima of CC measures can be obtained. However, we want to study  whether the algebraic maximum value of these quantities can also be reached for randomly generated states.

\subsection{Class of states maximizing classical correlation measures}
\label{sec:ccmax}

Before continuing our study with random states, let us  determine  the class of states for which all individual  two-party classical correlations in a multiqubit state simultaneously reach algebraically maximal values.  Specifically, we identify states which maximize \(\sum_{i = 2}^N C_{1i} \). For two-qubit states, since each \(C_{1i}\) can be unity, \(\sum_{i = 2}^N C_{1i} \) can, in principle, reach \(N-1\). For all the CC measures discussed above, it is indeed possible to saturate that bound for a certain types of states. To illustrate this, we consider product states,  $|0\rangle^{\otimes N}$ and $|1\rangle^{\otimes N}$ as well as Greenberger-Horne-Zeilinger (GHZ) state \cite{GHZst}, $ |GHZ\rangle = \frac{1}{\sqrt{2}}(|0\rangle^{\otimes N} + |1\rangle^{\otimes N})$, which also possess the maximal amount of genuine multiparty entanglement \cite{GGM}. Note that for the GHZ state, all bipartite reduced states with party $1$ as the nodal observer read as $\rho_{1i} = \frac12 |00\rangle\langle 00| + \frac12 |11\rangle \langle11|$, for $i \geq 2$, while all single party reductions are same which is the maximally mixed state, i.e., $\rho_j = \frac12 \mathbb{I}_2$ for all $j \in [1,N]$.


1. \emph{For the classical correlator(s)} $C^{zz}_{1i} = |\langle \sigma_z \otimes \sigma_z \rangle_{1i}| = |\text{tr } (\sigma_z \otimes \sigma_z \rho_{1i})| = 1$ for the \(|GHZ\rangle\) state.  Naturally, we also get the same results for the states $|0\rangle^{\otimes N}$ and $|1\rangle^{\otimes N}$.  Thus, we have $\sum_{i=2}^N C^{zz}_{1i}$ to be $N-1$, the algebraic maximum for all these three states. Let us now consider the covariance of
$C^{zz}_{1:i}$ given by $\tilde{C}^{zz}_{1:i} = |\langle \sigma_z \otimes \sigma_z \rangle_{1i} - \langle \sigma_z \rangle_1 \langle \sigma_z \rangle_i|$, 
for both $|0\rangle^{\otimes N}$ and $|1\rangle^{\otimes N}$. Now,  $\langle \sigma_z \otimes \sigma_z \rangle_{1i} = \langle \sigma_z \rangle_1 \langle \sigma_z \rangle_i = 1$. Therefore, we get $\tilde{C}^{zz}_{1i} = 0$, identically. On the contrary, since for the GHZ state, $\langle \sigma_z \rangle_1 = \langle \sigma_z \rangle_i = \frac12 \text{tr } (\sigma_z I) = 0$, we obtain $\tilde{C}^{zz}_{1i} (|GHZ\rangle) = 1$. Hence, we have $\sum_{i=2}^N \tilde{C}^{zz}_{1i} = N-1$ only for the GHZ state. Similar analysis can also be performed for other classical correlators and corresponding states can be identified. 

2. \emph{Classical part of QD (CQD).} 
Let us compute  CQD of \(\rho_{1i}\) for the \(|GHZ\rangle\) state.  When a measurement is performed on the second party (i.e., the $i$-th party) in the $\lbrace|0\rangle,|1\rangle \rbrace$-basis, we get pure post measurement states $|0\rangle$ and $|1\rangle$ with equal probabilities. Hence the second term of Eq. \eqref{eq_CCQD} vanishes, thereby maximizing the total quantity. Furthermore, the first term of Eq. \eqref{eq_CCQD} is unity since, as pointed earlier, all single party reduced density matrices are maximally mixed states. Therefore, $C^{D}_{1i}(|GHZ\rangle) = 1$, and consequently by summing over $i$, we get the algebraic maximal value.

3. \emph{Local work.} The second term in the definition of $C^{LW}_{12}$ in Eq. \eqref{eq_lwdef}, takes the minimum value of zero for any pure product state, $|\psi\rangle_1 \otimes |\psi\rangle_2$. Therefore,  for any $N$-qubit pure completely product states $\otimes_{k=1}^N |\psi\rangle_k$, $C^{LW}_{1i} = \frac12(2-0) = 1$. Consequently, we get $\sum_{i=2}^N C^{LW}_{1i} = N-1$. 

Note that although we can possibly provide additional examples that give the algebraic maximal values, we, unfortunately, cannot provide an exhaustive set of states with this property. We think this, in general, is a difficult problem. Furthermore, note that numerical analysis would not help since the probability of a randomly generated state to reach the algebraically maximal value is vanishingly small.

\emph{Remark:} In the Hilbert Schmidt representation of a two-qubit state, the Bloch coefficients $C^{kl}$, $k,l = x,y,z$, form the elements of the correlation matrix. Any one of the nine coefficients does not contain any “quantum” properties, since a separable state also can possess exactly the same value of the given Bloch coefficient. However, when one takes all the Bloch coefficients, i.e., the entire correlation matrix, of course, it has quantum properties and cannot be dubbed as merely classical. For example, the trace of the absolute values of the correlation matrix, $N = C^{xx} + C^{yy} + C^{zz}$ is directly connected to the average teleportation fidelity, $F = \frac12 (1 + N/3)$ \cite{telefid}. Therefore, this by no means can be considered as classical. So, what we understood was that taking multiple Bloch coefficients simultaneously is tricky when we want to exclusively probe classical properties. So we resort to the use of single Bloch coefficients as a measure of classical correlation. As mentioned before, using covariances, 
we find that the classical correlation of $|00\rangle$ is zero ($\tilde{C}^{zz}_{1i} = 0$).
However, surprisingly, both the variants of CC (single Bloch coefficients and the covariance-based ones) produced the same qualitative statistical features for Haar uniformly generated random states, as shown in Fig. \ref{fig:czzuncon}.  We present the results for the single Bloch coefficients for their wide applicability in various areas in quantum information science. 

Notice that we will keep these states out from our analysis since we are only concerned about properties of random multiqubit states. When states are chosen randomly, the probability that one picks states from these classes is vanishingly small. Hence,  a new upper bound lower than the algebraic maxima may emerge for almost all states (as sampled by Haar uniform generation \cite{Karolbook}), since all the measure zero states would naturally be eliminated from our analysis.
Next section  focuses on the possibility of any form of restriction on the distribution of classical correlations among bipartite reduced states of random multiparty quantum states. 

\section{Trends of shareability of classical correlations for unconstrained random states} 
\label{sec:uncon1}

We first generate  three-, four-, five- and six-qubit pure states   Haar uniformly  \cite{Karolbook} and compute their possible two-party  reduced density matrices shared between the nodal observer and other parties, i.e., in our case, \(\rho_{1i}, i=2,\ldots, N\) obtained from a pure state, \(|\psi\rangle_{12\ldots N} \). From these generated states, we estimate sum of their CCs  without imposing any additional condition on its properties and  perform the analysis for the classical correlators, the classical part of quantum discord, and the local work  of quantum work deficit. 

\subsection{Rule for distributing classical correlators in random multipartite states}

We begin by looking at the CCCs, $C^{\alpha \beta}$, where $\alpha,\beta \in \lbrace x, y, z \rbrace$, as defined in Eq. (\ref{eq_CCdef}). 
Our analysis reveals that all the $C^{\alpha \beta}$s display qualitatively and quantitatively similar features, and so without loss of generality, we focus on a particular one, say $C^{xx}$. 


\begin{figure}
\includegraphics[width=\linewidth]{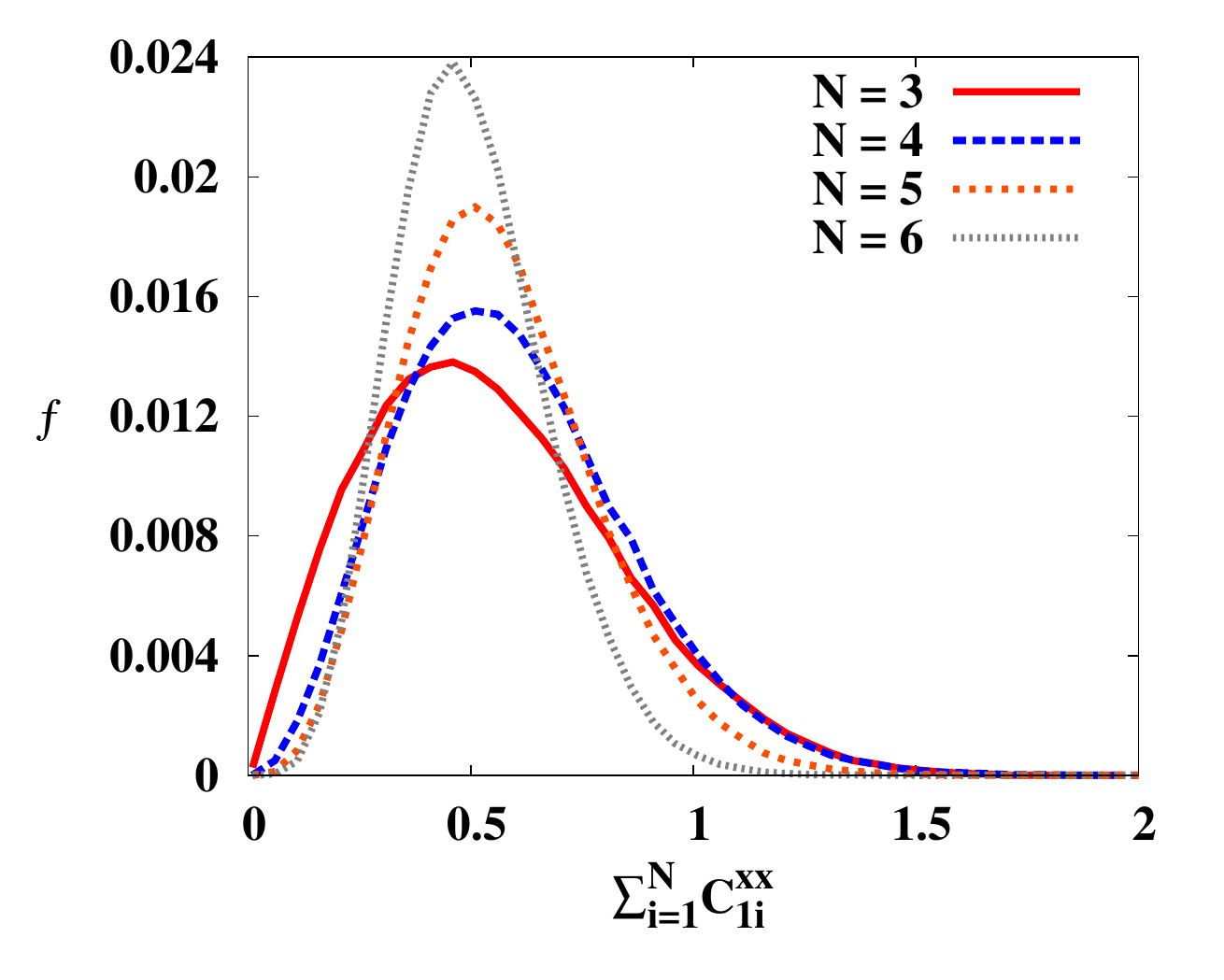}
\caption{(Color online.) Frequency distribution of   $\sum_i C^{xx}_{1i}$. The fraction of states, $f$, (vertical axis) is plotted against $\sum_i C^{xx}_{1i}$ (horizontal axis) with a bin size of $0.01$.
Total number of random states generated for the analysis for each $N$ is $10^6$. Note here that  the covariance of \(\sum_i C^{xx}_{1i}\) also gives the similar frequency distribution, having almost the same mean and standard deviation.   All the axes are dimensionless.}
\label{fig:czzuncon}
\end{figure}

\begin{table}
\begin{center}
\begin{tabular}{|c|c|c|c|c|c|}
\hline
$N$ &  3 & 4   & 5 & 6  \\
     \hline
mean & 0.546    & 0.589   & 0.559   & 0.497   \\     
\hline
sd & 0.281 &  0.258 & 0.214 & 0.170 \\ 
\hline
max val & 1.856  & 2.101 & 2.026 & 1.441 \\
\hline
\end{tabular}
\end{center}
\caption{Statistical data for the distribution of the fraction of states, \(f\), obtained for \(\sum_{i=2}^N C^{xx}_{1i}\). The mean, standard deviation, and the maximum value of \(\sum_{i=2}^N C^{xx}_{1i}\)for randomly generated states are denoted respectively by mean, sd, and max val. The total number of randomly generated state is \(10^6\).}
\label{table:cc-uncon}
\end{table}

Let us enumerate  below the observations of the distributions for \(\sum_i C^{xx}_{1i}\) as depicted in Fig. \ref{fig:czzuncon} and Table \ref{table:cc-uncon}: 
\begin{enumerate}
\item We trace out the fraction of randomly generated states, $f$, which possess $\sum C^{xx}_{1i}$ values in a range denoted by a step size of $0.01$ among \(10^6\) samples, i.e. \[f = \frac{\mbox{Number of states having values} \, \alpha < \sum C^{xx}_{1i} \leq \alpha + 0.01}{\mbox{Total number generated states}}, \] where \(\alpha\) is a fixed value of \(\sum_i C^{xx}_{1i}\), and $0.01$ is the bin size in this case which will be changed depending on the analysis.  We find that $f$ depicts ``Gaussian"-like features for  all chosen  number of qubits, i.e., $N \leq 6$. 

\item \emph{Mean and standard deviation.} With the increase of $N$, the standard deviation of the $f$-distribution decreases, thereby  making it more spiked as shown in Fig. \ref{fig:czzuncon}. However,  mean of the distribution remains almost constant with $N$  (see Table \ref{table:cc-uncon}).

\item \emph{Algebraic maximum.} For three-qubit random states, we can find  states for which   $\sum_{i = 2}^3 C_{1i}^{xx}$ is very close to its algebraic maximum, $2$. However, for larger $N$ values, maximal value obtained for $\sum_{i = 2}^N C_{1i}^{xx} $ is much lower compared to the algebraic maximum, \(N-1\), as can be compared also from the Table \ref{table:cc-uncon}.
\end{enumerate}

\begin{figure}[h]
\includegraphics[width=\linewidth]{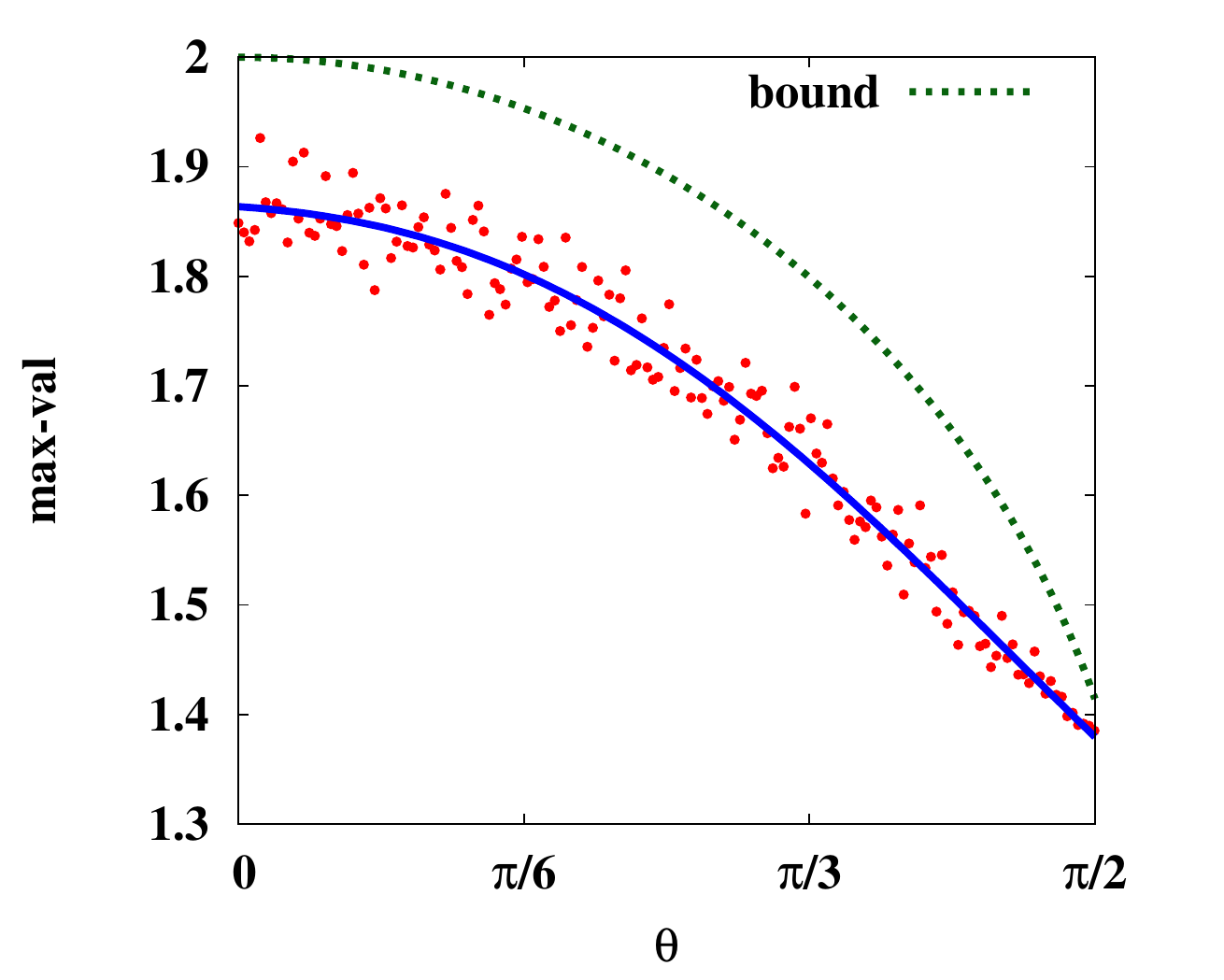}
\caption{(Color online.)
Maximum  of $C_{12}^{kx}+C_{13}^{xx}$ (\(y\)-axis) vs. \(\theta\)  (in \(x\)-axis) where $k$ is the unit vector, $(\cos \theta, \sin \theta, 0)$.  Red points represent maximal values of  $C_{12}^{kx}+C_{13}^{xx}$ for different $\theta$ and the blue line is the best  fit, depicting the decreasing nature with the increase of noncommutativity as measured by \(\theta\). Dotted line is the upper bound obtained in Eq. (\ref{eq_bound_new}). All the axes are dimensionless.   }
\label{fig:comm}
\end{figure}
\subsubsection{Role of observable incompatibility}
\label{sec:incomp}
\begin{figure*}[ht]
\includegraphics[width=0.8\linewidth]{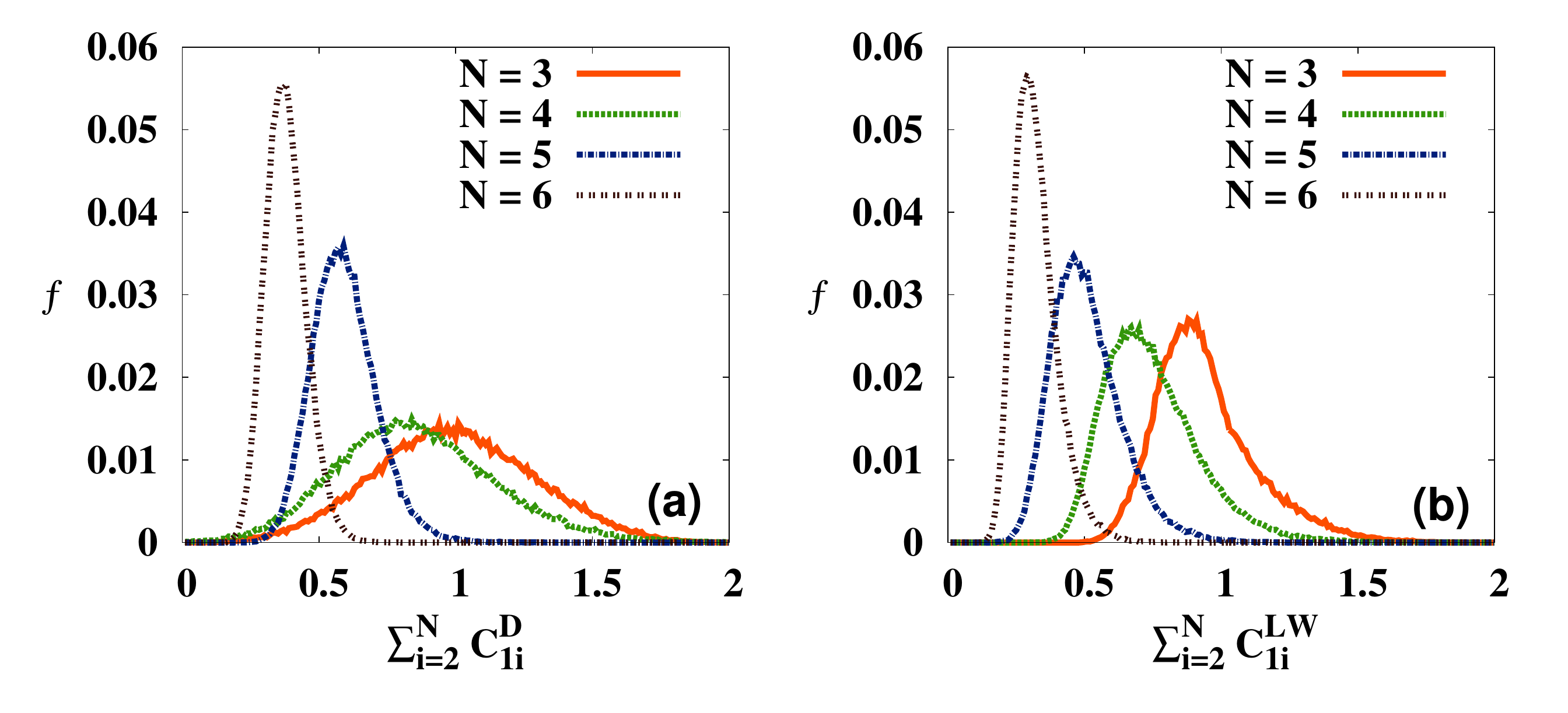}
\caption{(Color online.) Frequency distributions of  $\sum_i C^{D}_{1i}$ in (a), and $\sum_i C^{LW}_{1i}$  in (b) for \(N = 3\) to \(6\) parties.  The other specifications are same as in Fig. \ref{fig:czzuncon}. }
\label{fig:disruncon}
\end{figure*}

So far, all the  CCCs, $C_{1i}^{\alpha \beta}$,  involved in the sum $\sum_{i = 2}^N C_{1i}^{\alpha \beta} $  were the same, i.e., they possess the same $\alpha$ and $\beta$ values for all $i \geq 2$. However,  one may ask how the distribution changes if  $\alpha$ and $\beta$ change with $i$. In particular, it will be interesting to know how the distribution  of $f$ or the maximal value  changes when the classical correlators for different $i$ values do not commute. 

For illustration, in the three-party case, we consider $C_{12}^{yx}+C_{13}^{xx}$. We find that although the $f$-distribution does not vary much from   $C_{12}^{xx}+C_{13}^{xx}$, the maximal value of the sum of the correlators decreases as operators become more incompatible in the sense of noncommutativity.

Towards checking it, we now investigate how the maximal value varies on changing the commutativity of the operators i.e., when the operators become non-commuting from the commuting ones.  For a quantitative analysis, we compute the $f$-distribution for $C_{12}^{kx}+C_{13}^{xx}$, where the direction $k$ is defined by the unit vector, $(\cos \theta, \sin \theta, 0)$. The corresponding local operator for the direction $k$ is defined by $\cos \theta \sigma_x + \sin \theta \sigma_y$. Note that $\theta = 0$ represents the commuting case, while $\theta  = \pi/2$ refers to the maximum non-commuting ones. As $\theta$ increases, i.e, when the amount of incompatibility between two operators  increases, we observe that the maximal value of $C_{12}^{kx}+C_{13}^{xx}$ decreases, see Fig. \ref{fig:comm}. We now attempt to provide  an upper bound of $C_{12}^{kx}+C_{13}^{xx}$ following the ideas in \cite{com1, com2, com3}. For this, we first consider two projectors $A$ and $B$, and construct an operator $M = A\langle A\rangle + B \langle B \rangle$. We now have the following:
\begin{eqnarray}
\langle M \rangle &=& \langle A \rangle^2 + \langle B \rangle^2, \nonumber \\
\langle M^2 \rangle &=& \langle M \rangle + \langle A \rangle\langle B \rangle \langle \lbrace A,B \rbrace \rangle,
\end{eqnarray}
where $\lbrace A,B \rbrace = AB + BA$. Using the non-negativity of the variance $\langle M^2 \rangle - \langle M \rangle^2 \geq 0$ and approximating $\langle A \rangle \langle B \rangle = 1$, we arrive at the  relation, given by
\begin{eqnarray}
\langle M \rangle^2 - \langle M \rangle - \langle \lbrace A,B \rbrace \rangle \leq 0.
\end{eqnarray}
Since $\langle M \rangle \geq 0$, the above condition reduces to
\begin{eqnarray}
\langle M \rangle \leq \frac12 (1 + \sqrt{1 + 4 \langle \lbrace A,B \rbrace \rangle}  ).
\label{eq:comm1}
\end{eqnarray}
Again, using the Cauchy-Swartz inequality, we have
\begin{eqnarray}
|\langle A \rangle| + |\langle B \rangle| \leq \sqrt{2}\sqrt{\langle A \rangle^2 + \langle B \rangle^2} = \sqrt{2}\sqrt{\langle M \rangle}.
\end{eqnarray}
Using Eq. \eqref{eq:comm1}, we finally obtain 
\begin{eqnarray}
|\langle A \rangle| + |\langle B \rangle| \leq 1 + \sqrt{1 + 4 \langle \lbrace A,B \rbrace \rangle}.
\label{eq:comm2}
\end{eqnarray}
If we now put $A = (\cos \theta\sigma_x + \sin \theta \sigma_y) \otimes \sigma_x \otimes \mathbb{I}_2$ and $B = \sigma_x \otimes \mathbb{I}_2 \otimes \sigma_x$, we get  
\begin{eqnarray}
\langle \lbrace A,B \rbrace \rangle &=& 2\cos \theta \langle\mathbb{I}_2 \otimes \sigma_x \otimes \sigma_x \rangle + \nonumber \\
  \sin \theta \langle(\{ \sigma_x &\otimes& \sigma_y \}) \otimes \sigma_x \otimes \sigma_x \rangle \leq 2\cos \theta.
\end{eqnarray}
Furthermore, the above substitution makes $|\langle A \rangle | = C_{12}^{kx}$ and $|\langle B \rangle | = C_{13}^{xx}$. Finally, pulling everything together, Eq. \eqref{eq:comm2} becomes
\begin{eqnarray}
C_{12}^{kx} + C_{13}^{xx} \leq \sqrt{1 + \sqrt{1+8 \cos \theta}}.
\label{eq_bound_new}
\end{eqnarray}
To test the quality of this bound, we plot it in Fig. \ref{fig:comm} along with our numerical findings. The above bound turns out to be good as can be clearly seen in Fig. \ref{fig:comm}.

Although the mean of the frequency distribution is independent of $\theta$, the reduction in maximal value is due to the lowering of the standard deviation of the distribution induced by increasing incompatibility. The behavior obtained above remains qualitatively similar  for any two noncommuting operators, say $C_{12}^{k_1 k_2} $  and $C_{13}^{l_1 l_2} $ in the sum while the maximal value remains same for two commuting operators in $\sum_{i = 2}^N C_{1i}^{\alpha \beta} $. For example, we find that the maximum of  $C_{12}^{xk}+C_{13}^{xx}$ matches with that of  $C_{12}^{xx}+C_{13}^{xx}$ since  $C_{12}^{xk}$ commutes with $C_{13}^{xx}$. 
 This further reinforce that the reduction of the maximal value is due to the incompatibility of operators. Such reduction of maximal values for incompatible operators is observed for higher  $N$-values as well.

\subsection{Equivalent shareability rule for classical  part of discord and local work  }

\begin{table}
\begin{center}
\begin{tabular}{|c|c|c|c|c|}
\hline
$N$ &  3 & 4   & 5 & 6 \\
     \hline
mean &  0.989 & 0.848 & 0.587 & 0.373  \\     
\hline
sd & 0.291 & 0.289 & 0.117 & 0.073  \\ 
\hline
max val & 1.946 & 2.207 & 1.337 & 0.925  \\
\hline
\end{tabular}
\end{center}
\caption{
Mean and standard deviation of the frequency distribution are obtained for $\sum_i C^{D}_{1i}$ with different values of $N\leq 6$. The maximum of the sum is also computed by increasing $N$. The other specifications  are same as in Table \ref{table:cc-uncon} with the exception that the number of states sampled for each $N$ is $10^5$.}
\label{table:cd-uncon}
\end{table}

\begin{table}
\begin{center}
\begin{tabular}{|c|c|c|c|c|}
\hline
$N$ &  3 & 4   & 5 & 6  \\
     \hline
mean &  0.937  & 0.741 & 0.503 & 0.316   \\     
\hline
sd & 0.183 & 0.172 & 0.128 & 0.079  \\ 
\hline
max val & 1.877 & 1.962 & 1.380 & 0.883  \\
\hline
\end{tabular}
\end{center}
\caption{(Color online.) Mean and standard deviation of $f$ for  $\sum_i C^{LW}_{1i}$ and the maximum of it are tabulated. For  other specifications, see Table \ref{table:cd-uncon}.}
\label{table:cdef-uncon}
\end{table}

We now concentrate our analysis on the classical part of QD and local work. As we have argued, the classical correlators have a completely different origin than the CQD and LW and hence we may expect some  qualitative differences between classical correlators and CQD or  LW  with the increase of $N$.   Finally, we also compare the trends of $f$ obtained for CQD and LW. 

\begin{figure}[h]
\includegraphics[width=\linewidth]{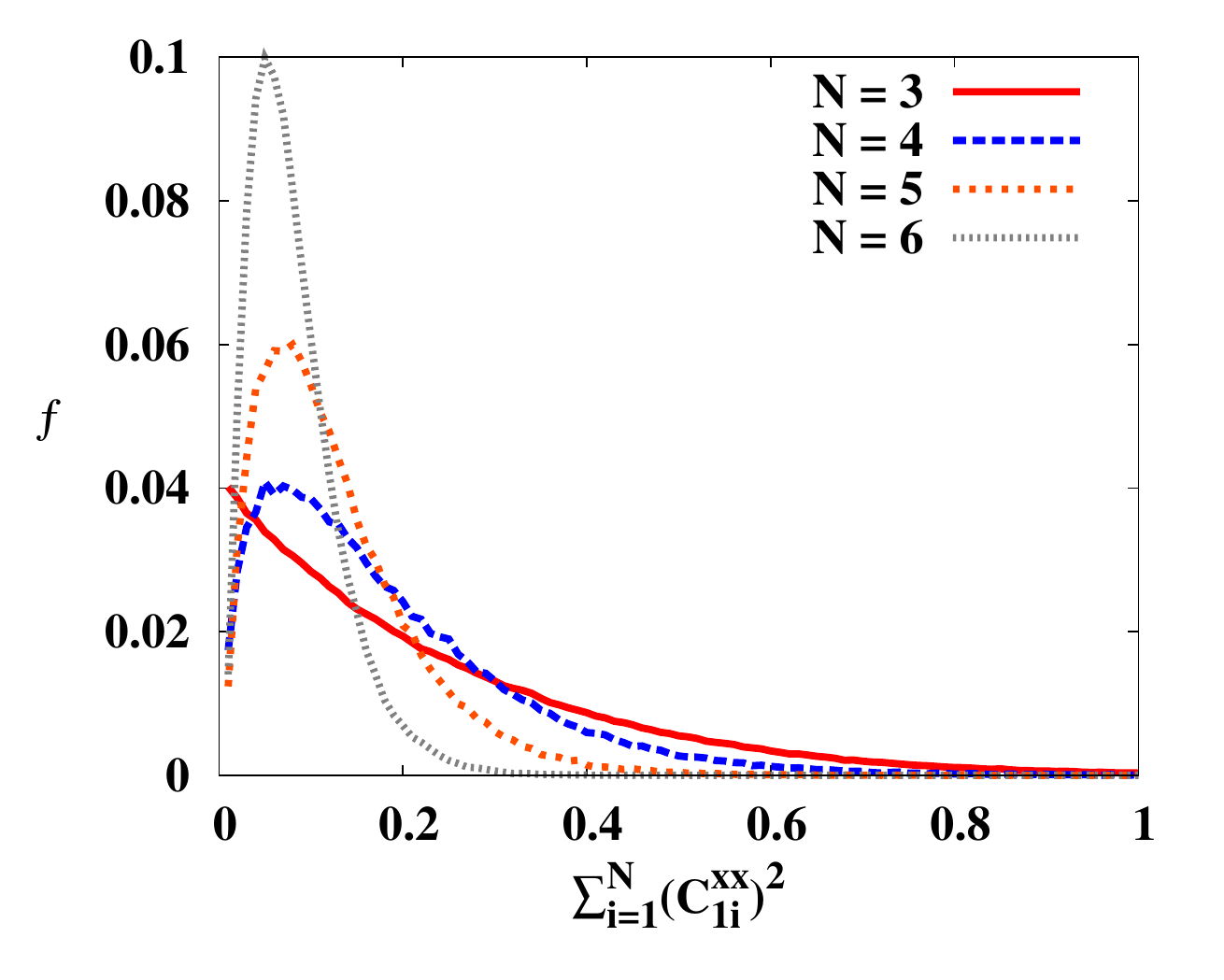}
\caption{(Color online.)
Frequency distribution of   $\sum_{i=2}^N (C^{xx}_{1i})^2$. The fraction of states, $f$, (vertical axis) is plotted against $\sum_{i=2}^N (C^{xx}_{1i})^2$ (horizontal axis) with a bin size of $0.01$.
Total number of random states generated for the analysis for each $N$ is $10^5$. All the axes are dimensionless.}
\label{fig:ccc2}
\end{figure}

\begin{figure}[h]
\includegraphics[width=\linewidth]{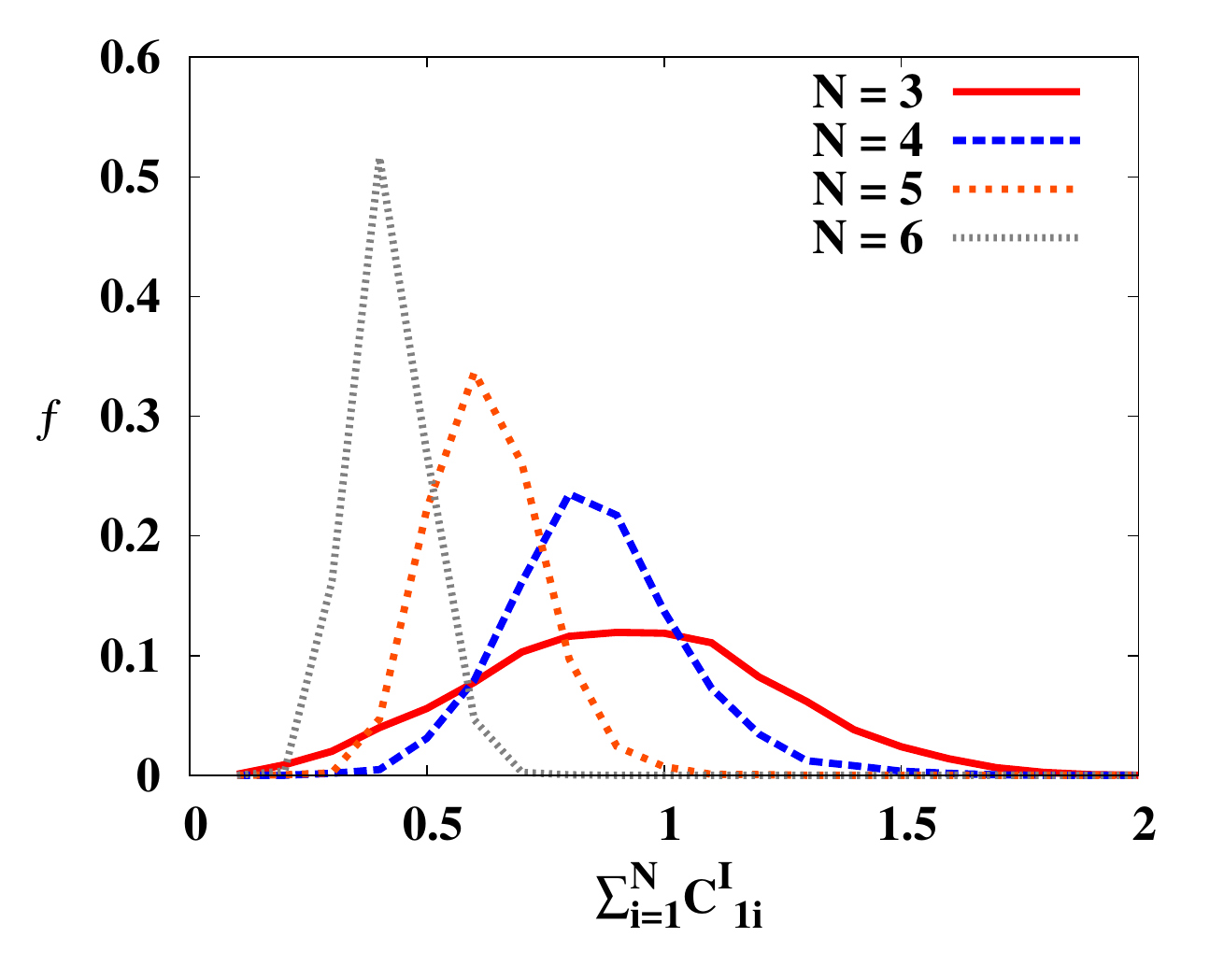}
\caption{(Color online.)
Frequency distribution (f) of   $\sum_{i=2}^N C^{I}_{1i}$ (vertical axis) vs. $\sum_{i=2}^N C^{I}_{1i}$ (horizontal axis) as defined in Eq. (\ref{eq_maxmutual}). All other specifications  are same as in Fig. \ref{fig:ccc2}. }
\label{fig:mmi}
\end{figure}

\begin{figure*}[ht]
\includegraphics[width=0.8\linewidth]{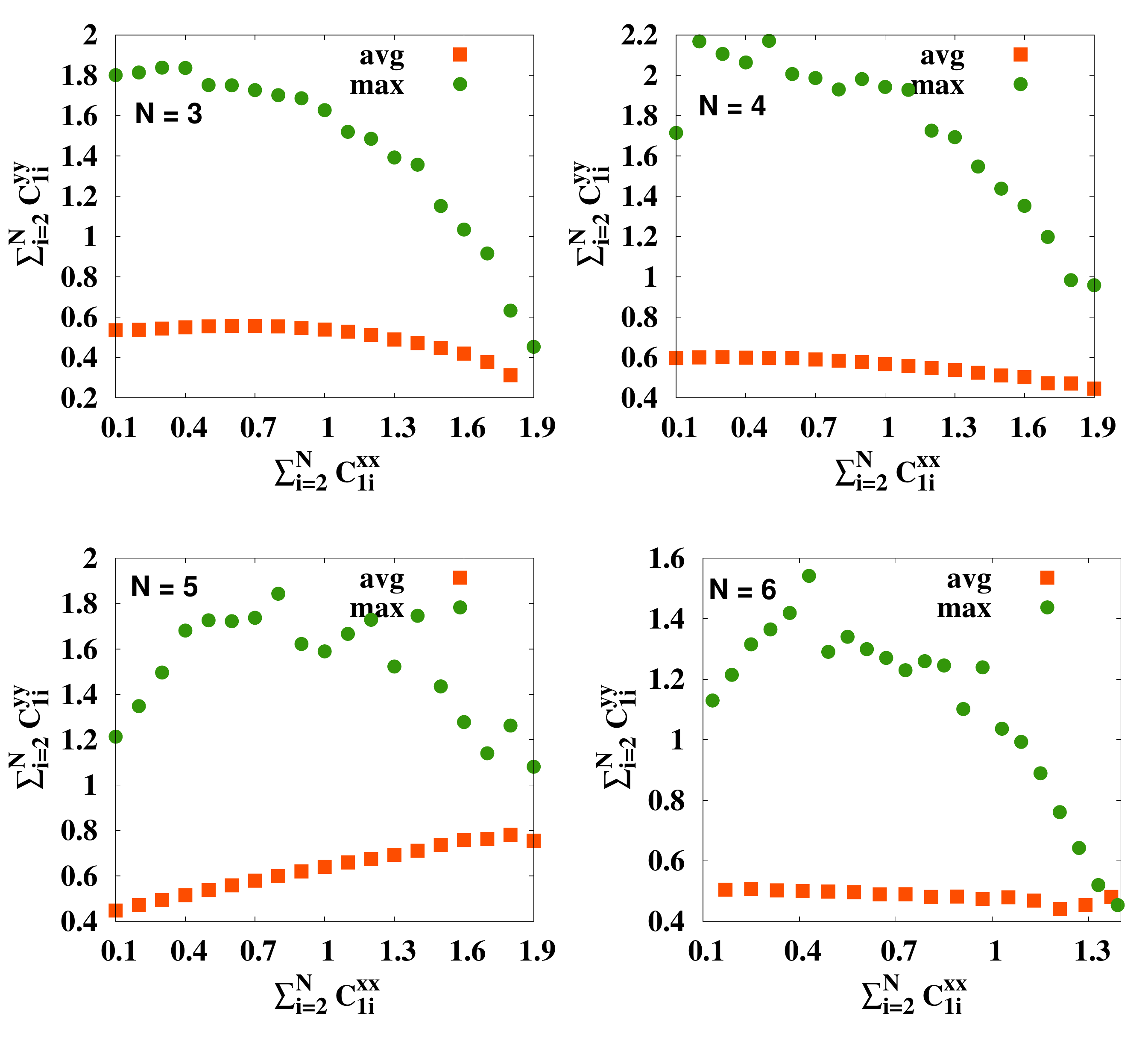}
\caption{(Color online.) Average (avg) and maximal values (max) of $\sum_{i=2}^N C_{1i}^{yy}$ (ordinate) for random \(N\)-party states possessing  a definite amount of  $\sum_{i=2}^N C_{1i}^{xx}$ (abscissa). Squares represent the average values while circles are for maximum value. 
$N = 3$ and $4$ are in upper panel while in lower panel from left,  $N=5$ and $6$ are displayed. Here the bin size is set to $0.1$. All axes are dimensionless.}
\label{fig:cxx-cyy}
\end{figure*}

%
The statistical analysis leads to the emergence of some important features which we now list down below:
\begin{enumerate}

\item \emph{f-distribution.} The shapes of the frequency distribution of \(\sum_{i=2}^N C^{D}_{1i}\) and \(\sum_{i=2}^N C^{LW}_{1i}\)  for \(N\leq 6\) are similar to the one obtained from classical correlators, as seen by comparing Figs. \ref{fig:czzuncon} and \ref{fig:disruncon}.

\item \emph{Mean from CQD and LW.} Unlike the classical correlators, for which the mean of the $f$-distribution remains almost invariant on changing $N$, the mean of the $f$-distribution for the CQD and LW decreases monotonically on increasing $N$ (compare Tables  \ref{table:cd-uncon} and \ref{table:cdef-uncon}). 
Surprisingly,  we find that means of the distribution  obtained from CQD and LW behave even quantitatively similarly. 

\item \emph{Standard deviation of the distribution from CQD and LW.} Like the classical correlators, the standard deviation of the distribution  decreases progressively on increasing $N$.

\item \emph{Algebraic maxima.} The maximal value of   \(\sum_{i=2}^N C^{D}_{1i}\) and \(\sum_{i=2}^N C^{LW}_{1i}\) decreases sharply on increasing $N$. This prompts us to think whether we can put an upper bound to the sum for random multiqubit pure states. The question will be addressed in the subsequent sections.
\end{enumerate}

Interestingly, note that the trends of the frequency distributions for classical correlators  are quite different from that of the classical part of quantum  discord and local work while the similarities in the distributions are observed for CQD and LW even when they are defined from two disjoint notions.  It might be worthwhile to investigate whether obeying (or disobeying) the postulates of classical correlations has some bearing on the differences or similarities in the observed features.


\subsection{Features for other measures of CC}
Apart from the CC measures discussed above, we also consider two other measures of CC. The first one is just the squares of the  classical correlators,  $(C^{kl})^2$ instead of taking their absolute values to ensure their non-negativity. Taking it as a measure of CC, we find  that it possesses qualitatively similar features as obtained by  using the absolute values. However, in this case, the frequency distribution is highly skewed to the left (see Fig. \ref{fig:ccc2}), especially for low number of parties. This is due to the fact that squaring has actually made the correlation values smaller since they are already (typically) less than unity. It also explains the reason behind the mean values of the frequency distribution to be  smaller compared to the case with absolute values of the correlations. When the absolute values are considered, it does not suffer from unnecessary value reduction, and hence supports our choice of considering absolute values to scale the values of the quantity from $0$ to $1$.

The second one is the maximal mutual information between local measurement results performed on a two-party state $\rho$, and is defined as follows:
\begin{eqnarray}
C^I = \max_{\Pi_a, \Pi_b} I(\Pi_a \otimes \Pi_b ~\rho),
\label{eq_maxmutual}
\end{eqnarray}
where $I(AB) = H(A) + H(B) - H(AB)$ denotes the mutual information content of the measurement statistics with \(H(.) = \sum_i p_i \log_2 p_i\) being the Shannon entropy. We now compute the frequency distribution of $\sum_{i=2}^N C^I_{1i}$ and   by comparing Figs. \ref{fig:czzuncon} and \ref{fig:mmi}, find that the statistical properties to be almost identical to those obtained by other CCCs. 

\begin{figure*}[ht]
\includegraphics[width=0.8\linewidth]{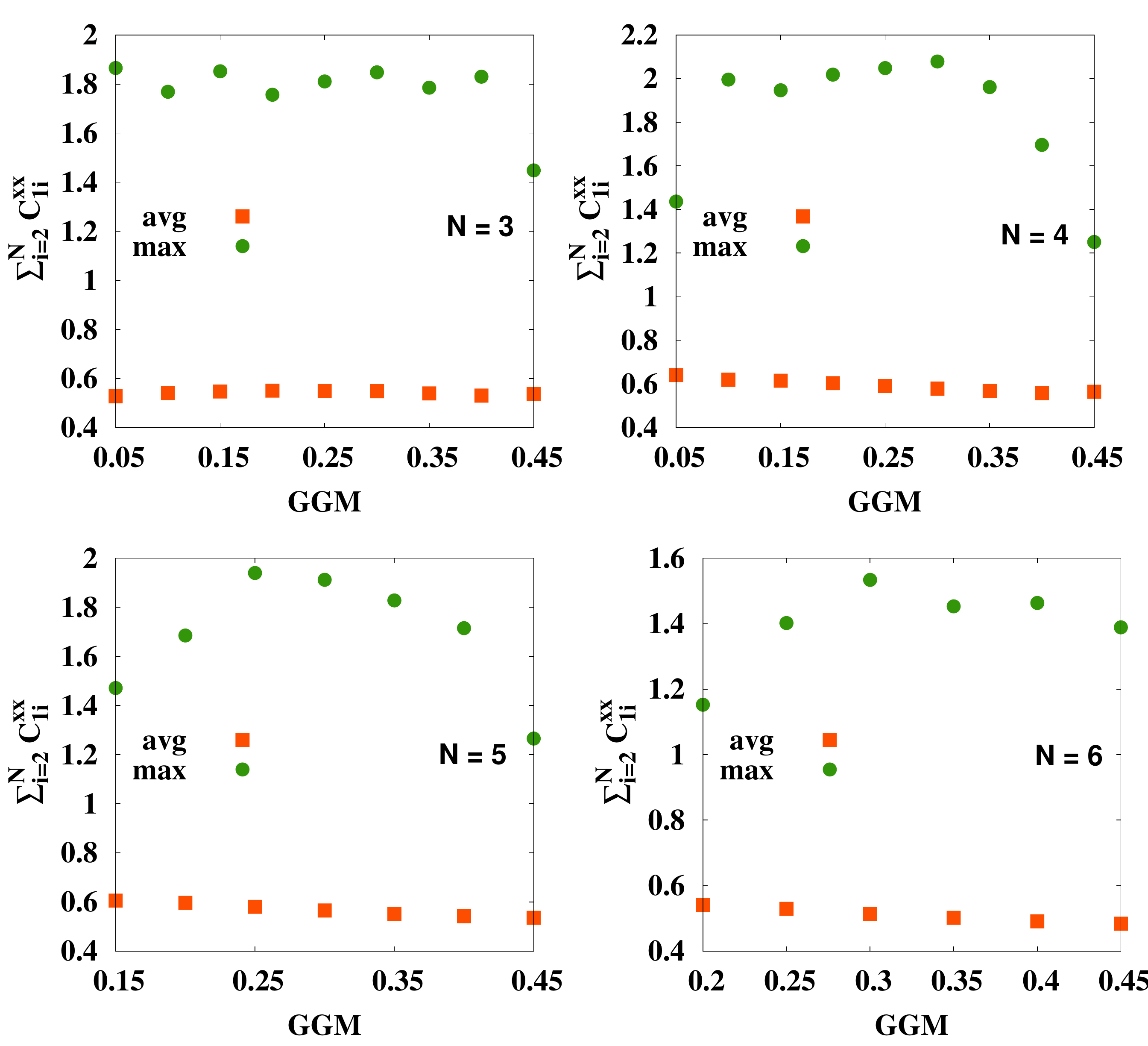}
\caption{(Color online.) Plot of an average and maximum of $\sum_{i=2}^N C_{1i}^{yy}$ (in \(y\)-axis) for random states having a fixed genuine multiparty entanglement content  as measured by GGM (in \(x\)-axis). Other specifications are same as in Fig. \ref{fig:cxx-cyy}, with the exception that the bin size in this case is $0.05$.  All axes are dimensionless.}
\label{fig:cxx-ggm}
\end{figure*}

\section{Distributions  of classical correlations for constrained random states} 
\label{sec:con1}

Let us now move to the investigations of the shareability of classical correlations for  randomly generated  multipartite states when  a fixed amount of a  particular physical property that can be both classical or quantum is available.
Moreover, we examine how the maximal values of  the CC measures can depend on the constraints, i.e. the choice and the range of the physical quantity of the random states. Like before, we perform our analysis for $3 \leq N \leq 6$.

\subsection{CCCs under constraints}

We now impose constraints either by fixing the range of the sum of bipartite CCC in transverse direction or, by  fixing the content  of the genuine multiparty entanglement  \cite{geoent, GGM}  of the randomly generated states. The latter can also answer  the role of classical correlators on a multipartite entanglement measure. 

\begin{figure*}[ht]
\includegraphics[width=0.8\linewidth]{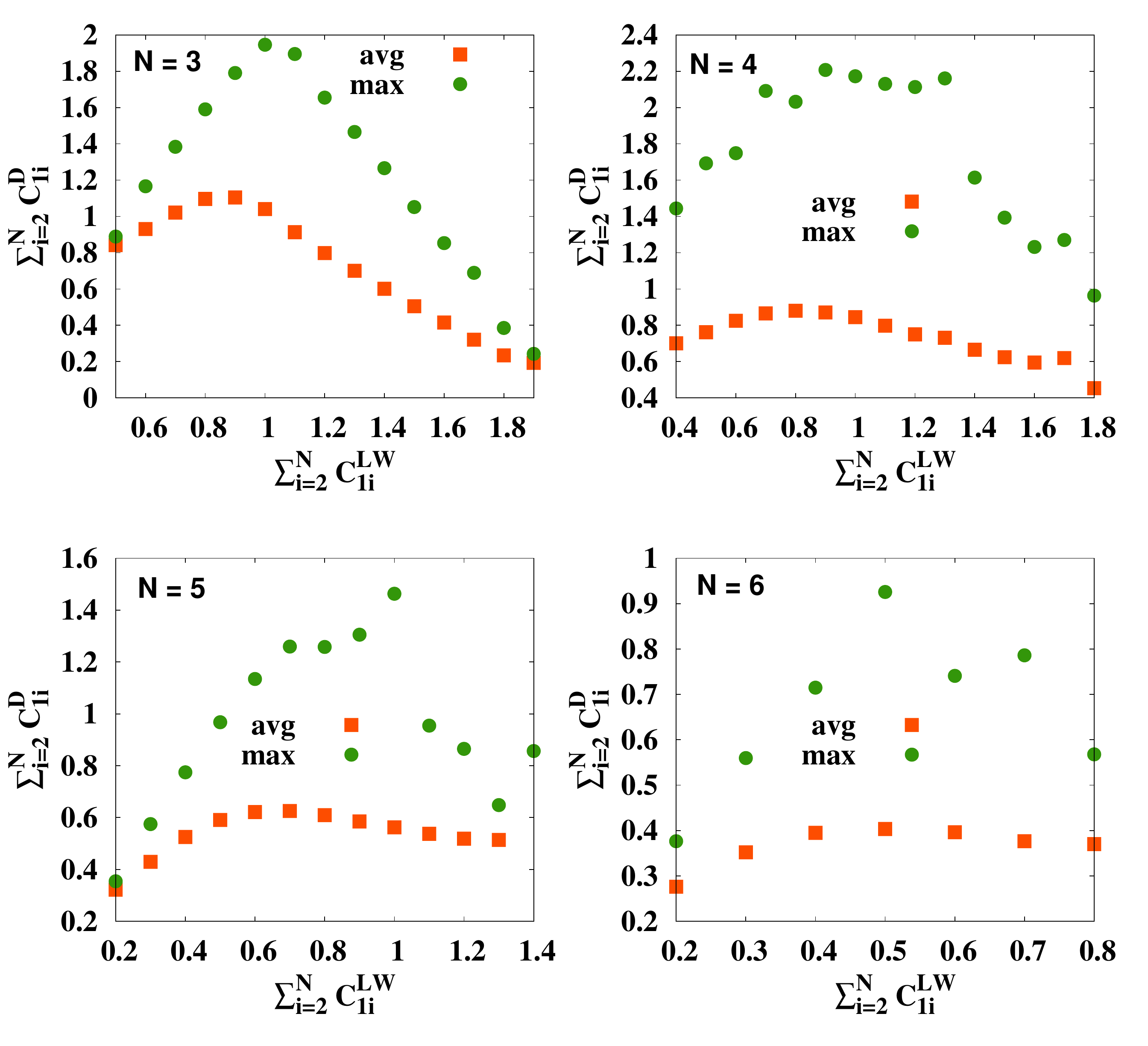}
\caption{ (Color online.) For a given amount of $\sum_i^N C^{LW}_{1i}$ (horizontal axis), average and maximal values of $\sum_{i=2}^N C^{D}_{1i}$ (vertical axis)  for random states is plotted for different values of \(N \leq 6\).  For other specifications, see  Fig. \ref{fig:cxx-cyy}. }
\label{fig:defdis}
\end{figure*}

\emph{Fixed ranges of CCC. } Let us first  reveal how restrictions on classical correlators in a fixed direction  effect the distribution of correlators in the transverse direction for multiqubit random states.  Without loss of generality,  we choose to study the distribution of \(C^{yy}\) for a fixed values of \(C^{xx}\).  In particular, 
we consider how the average and maximum  value of $\sum_{i=2}^N C_{1i}^{yy}$ depends on  a given amount  of $\sum_{i=2}^N C_{1i}^{xx}$ possessed by the random pure states. We lay out our findings below:

\begin{enumerate}
\item \emph{Three-party states.} For $N = 3$, we find that both the maximum and average of $C_{12}^{yy}+C_{13}^{yy}$ decreases with the increase of a quantity,  $C_{12}^{xx}+C_{13}^{xx}$, see Fig. \ref{fig:cxx-cyy} (a). It suggests that sum of bipartite classical correlators in transverse directions play a complementary role as  confirmed by the behaviors of both  average and maximal values. Similar feature is observed for $N = 4$.  It is important to note that such a dual behavior can also be seen if we choose any two noncommuting  classical correlators. This feature can be also viewed as a consequence of ``correlation complimentarity" as analyzed in Sec. \ref{sec:incomp}.

\item \emph{Higher number of parties.} On the contrary, a qualitatively different behavior  is observed when $N \geq 5$, specifically, we observe that  when the sum of the bipartite correlators in a particular direction grows, average of the sum of bipartite correlators  in the transverse direction remains almost constant, see Fig. \ref{fig:cxx-cyy} (c) and (d). 
Note that the maximal value of \(\sum_i C_{1i}^{yy}\) also  shows an initial increase with the increase of  \(\sum_i C_{1i}^{xx}\) but then displays an opposite behavior. 
\end{enumerate}

The above results reveal that unlike the unconstrained case, the features of these classical correlators in this constrained scenario strongly depend on the number of qubits of the sampled random states. For $N = 3$, when the maximal value is close to the algebraic maximum, we get a strong ``complementarity-type" behaviour while   a completely different  picture emerges with higher values of $N$.  Such an absence of  complementarity relation between CCCs in transverse directions for random states can be a consequence of the fact that  the gap between the allowed maximal value of \(\sum_{i} C_{1i}^{kk}\) and the algebraic maximum value for random states increases with $N$ and at the same time, the standard deviation decreases (see Table \ref{table:cc-uncon}).  

\emph{Fixed ranges of GGM.} 
Let us now consider the random states which  are segregated based on their genuine multiparty entanglement content (as measured by generalized geometric measure \cite{GGM, GGMdef} ). Specifically, we compute $\sum_i C^{xx}_{1i}$ for all the  random states having GGM values between say, \(\alpha\) and \(\beta\), where \(\alpha\) and \(\beta\) are fixed by the bin values,  i.e., \(\beta -\alpha =0.01\) in our case and  finally, we compute the average as well as the maximum of  $\sum_i C^{xx}_{1i}$. 
Note here that among Haar uniformly generated states, mean of GGM goes towards its maximum value with the increase of number of parties \cite{Eisertrand, Winterrand, Sooryarand, Ratulrand}. It implies that the bipartite content of entanglement decreases with $N$. On the other hand,  the observations for the distributions of bipartite classical correlators in random multipartite states   are as follows (see Fig. \ref{fig:cxx-ggm}):

\begin{enumerate}
\item  We find that the average value of $\sum_i C^{xx}_{1i}$ is almost independent of the GGM content  of sampled random states. In this respect, notice that the average value remains almost constant also for the unconstrained case, see Table. \ref{table:cc-uncon}. The feature of the constancy of the average value is independent of the number of qubits, $N$. It is also important to stress that although mean of multipartite entanglement increases with $N$, and hence $\sum_i E(\rho_{1i})$ decrease with $E$ being any entanglement measure, the effects of such behaviour cannot be captured only by $\sum_i C_{1i}^{xx}$.

\item Unlike the average values,  the maximal value of $\sum_i C^{xx}_{1i}$ for a fixed GGM  does not follow any strict pattern. However, it also does not change considerably with the GGM values of the sampled random states. 

\end{enumerate}
We will contrast this behaviour with that obtained for the other CC measures  considered in this paper in subsequent sections.

\begin{figure*}[ht]
\includegraphics[width=\linewidth]{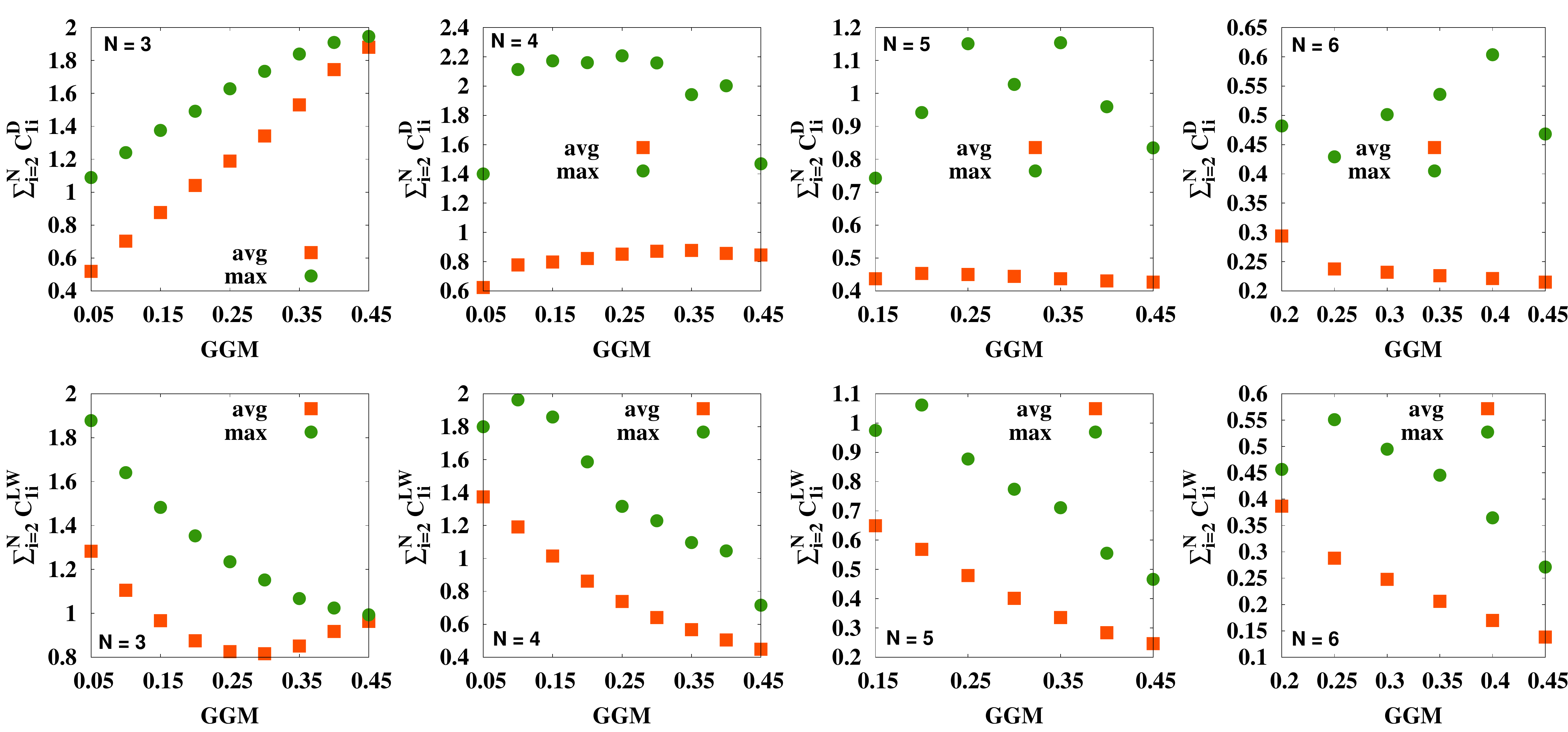}
\caption{(Color online.) Upper Panel:  Maximum and average values of  $\sum_{i=2}^N C^{D}_{1i}$ (ordinate) vs. GGM (abscissa). From left to right, $N$ increases from $3$ to $6$. 
Lower Panel:  $\sum_{i=2}^N C^{LW}_{1i}$  against GGM. Here bin size for computation is  used as $0.05$. All the axes are dimensionless. }
\label{fig:ggmdisdef}
\end{figure*} 

\subsection{CQD and LW for a fixed QC}   


\emph{Classical discord for a fixed content of  local work.}
Let us fix the sum of the amount of local work from various bipartite cuts of a multiparty state, i.e. when the value $\sum_{i=2}^N C^{LW}_{1i}$ lies between \(\alpha\) and \(\beta\) with \(\beta - \alpha\) being taken as \(0.01\), we find out the average and the maximal value of $\sum_{i=2}^N C^{D}_{1i}$. 

Our analysis reveals an emergence of a universal feature independent of the total number of qubits $N$. 
\begin{enumerate}
\item \emph{Average of CQD with LW constraints.}   For a fixed amount of  $\sum_{i=2}^N C^{LW}_{1i}$, we observe that the average of $\sum_{i=2}^N C^{D}_{1i}$  remains almost constant for high $N$. The change in average can only be seen with $N=3$ as shown in Fig. \ref{fig:defdis}.

\item \emph{Maximum under constraints.} The pattern of $\max \sum_{i=2}^N C^{D}_{1i}$ with respect to $\sum_{i=2}^N C^{LW}_{1i}$ is more drastic as compared to the average of the distribution. The pattern   can be  divided into two parts -- for  low values of  $\sum_{i=2}^N C^{LW}_{1i}$  ($\lessapprox 1$),  $\max \sum_{i=2}^N C^{D}_{1i}$ increases with the increase of  $\sum_{i=2}^N C^{LW}_{1i}$ while     interestingly,  a ``complementarity-type" relation  emerges when    
$\sum_{i=2}^N C^{LW}_{1i} \gtrapprox 1$. Specifically, in a latter case,  we get a decrease in $\max \sum_{i=2}^N C^{D}_{1i}$ values which ultimately become vanishingly small when the sum of local works goes close to its maximal values, see Fig. \ref{fig:defdis}. Such a behavior can  also be understood from the examples illustrated in Sec. \ref{sec:ccmax} and when  $\max \sum_{i=2}^N C^{D}_{1i}$ and $\max \sum_{i=2}^N C^{LW}_{1i}$ are studied for a given value of multipartite entanglement. 

\end{enumerate}

\begin{figure*}[ht]
\includegraphics[width=\linewidth]{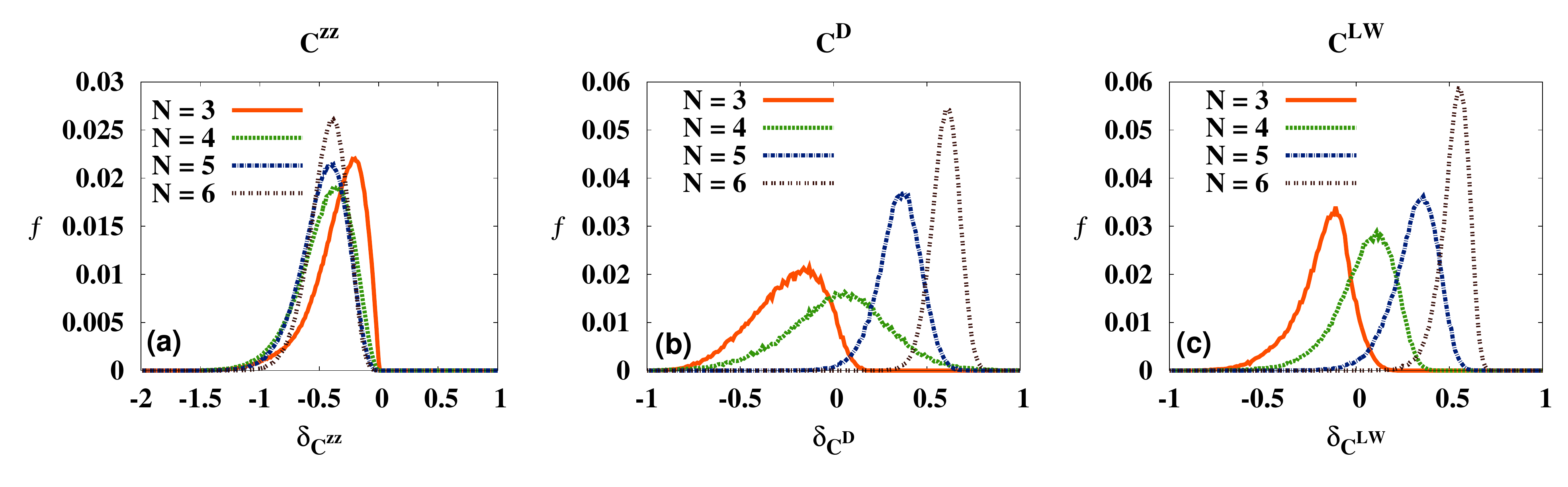}
\caption{(Color online.) Monogamy-motivated upper bounds from CC measures. From left to right: the frequency distribution, \(f\), in vertical axis is plotted for   \(\delta_{C^{zz}}\),  \(\delta_{C^D}\) and  \(\delta_{C^{LW}}\). \(N\) also varies from \(3\) to \(6\). All the axes are dimensionless. }
\label{fig:mscc1} 
\end{figure*}

\emph{Fixed  multipartite entanglement reveals dual nature of CQD and LW.}
For a given amount of  GGM in random three-, four-,  five and six-qubit states,  we observe a dual pattern in the maximum values for bipartite distributions of classical discord and  local work especially for $N = 3$ (see Fig. \ref{fig:ggmdisdef}). In particular, the maximal values of $\sum_{i=2}^N C^{D}_{1i}$ increase monotonically with increasing values of GGM, while we get the opposite feature for  $\sum_{i=2}^N C^{LW}_{1i}$.
Maximum of $\sum C_{1i}^{LW}$ always decreases with the increase of GGM. Let us now move to the average values of $\sum _i C_{1i}^{D/LW}$ with GGM. For $N \geq 4$, $\sum_i C_{1i}^{LW}$ always decreases while $\sum_i C_{1i}^{D}$ remains almost constant to a low value with the increase of GGM. As mentioned earlier for random states, it is known that mean GGM increases with $N$ and therefore one may expect low bipartite entanglement with increase in $N$. We find that $\sum_i C_{1i}^{D/LW}$ also follow the same trend as one may expect for bipartite entanglement. 
 Moreover, comparing Figs, \ref{fig:cxx-ggm} and \ref{fig:ggmdisdef}, it can again be established that the distributions of CCC among subsystems of random multipartite states are  quite distinct compared to that of the CQD and LW.  
 
 Remark: The properties when the constrained case is reanalyzed using $C^I$ (maximal mutual information between local measurement results) remains almost identical to the statistical features obtained for the CCCs.

\section{Bounding classical correlations}
\label{sec_boundingbymonogamy}

 As shown in Sec. \ref{sec:ccmax}, there always exists a quantum state for which the sum of bipartite classical correlations $\sum_{i=2}^N C_{1i}$ reaches the sum of the maximum of individual classical correlations. However, the results obtained in Secs \ref{sec:uncon1} and \ref{sec:con1} for Haar uniformly generated states strongly suggest that the measure zero subset of states possibly possesses the algebraic maximum value and therefore, for almost all states of the state space,  $\sum_{i=2}^N C_{1i}$  can be bounded by a smaller value than the algebraic maximum. Moreover, we observe that with increase in the number of parties, maximal values for all the classical correlation measures decrease and the gap between the algebraic maxima and the maxima for random states increases. 

Here we want to focus again on the upper bound of CC measures, motivated from the concept of monogamy of quantum correlations.
It is clear from the examples presented in Sec. \ref{sec:ccmax} that CC,  in general,  do not satisfy monogamy relation, thereby making it different from QC measure.   However, we intend to take a much more closer look at it for random states, since the results indicate that  for high values of $N$, the upper bound, $C_{1:\text{rest}}$, on the shareability of CC measure, may not be a bad bound for randomly generated quantum states. 
In particular, we construct a score for classical correlations as well, purely via a formal analogy,  examine the distribution of monogamy scores for any classical correlation measure, $C$, given by 
\(\delta_{C} = C_{1:\text{rest}} - \sum_{i=2}^N C_{1i}\) and 
 track the percentage of random states that do not satisfy the constructed monogamy relation.  

\begin{figure*}[ht]
\includegraphics[width=0.8\linewidth]{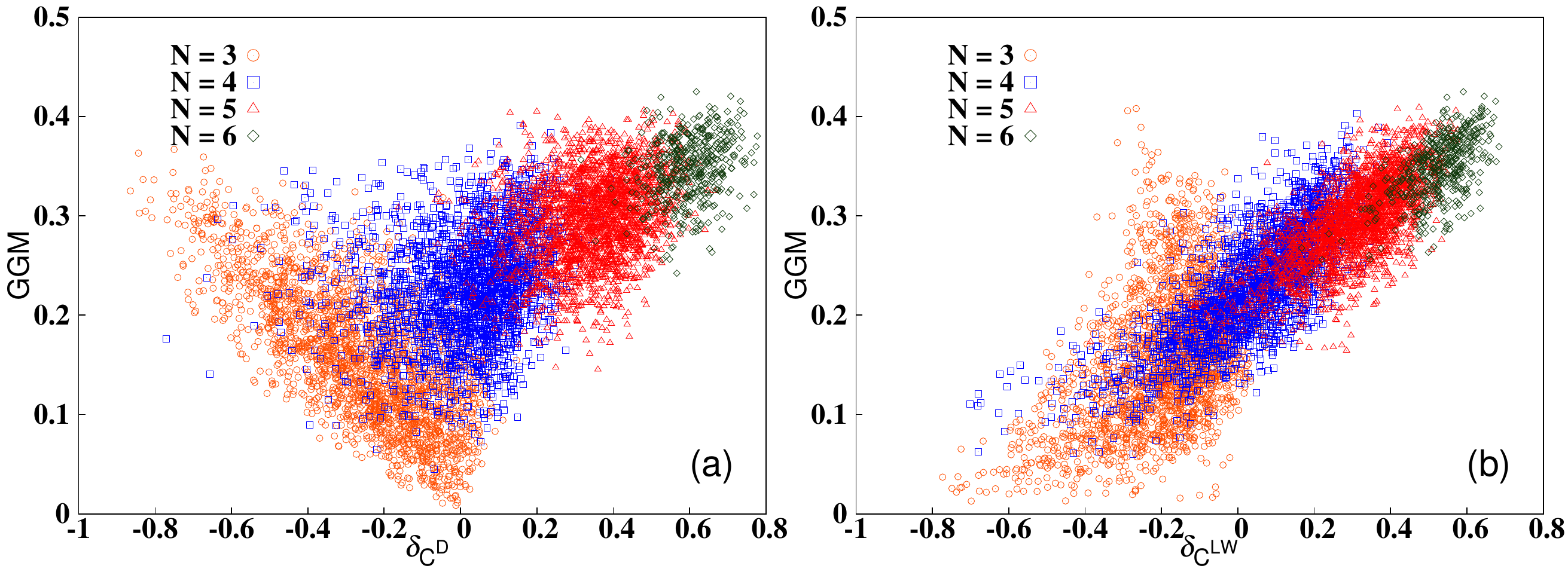}
\caption{(Color online.) Scatter plot for comparing \(\delta_{C^D}\) and  \(\delta_{C^{LW}}\) with GGM. We find that with the increase of $N$, higher GGM values (ordinate)  reveals monogamous nature of   \(\delta_{C^D}\) (abscissa in left panel) and  \(\delta_{C^{LW}}\) (abscissa in right panel). 
Circles, squares, triangles and diamonds correspond to \(N=3, 4, 5\) and  \(6\). 
All the axes are dimensionless.   }
\label{fig:mscc} 
\end{figure*}

\subsection{Monogamy-based upper bound for classical correlators}

As the prototypical classical correlator,  we take  $C^{zz}$. Firstly, note that for $C^{zz}_{1:\text{rest}}$, the ``rest" defines an $N-1$ qubit state formed by the parties, $2, 3, \ldots, N$. Therefore, the second $z$ in the superscript of $C^{zz}_{1:\text{rest}}$ represents spin $z$ operator for the $2^{N-1}$ dimensional system which in turn  corresponds to a spin of $s=\frac{2^{N-1}-1}{2}$. For spin-$s$, the magnetization along $z$-direction is measured by $\Lambda^z(s)$ whose matrix elements in the computational basis are given by
\begin{eqnarray}
[\Lambda^z(s)]_{ij} = 2(s-i)\delta_{ij} = 2(s-j)\delta_{ij}, 
\end{eqnarray}
where $0 \leq i,j \leq 2s$. It defines a diagonal matrix with entries diag$ \{ 2s, 2s - 2, 2s - 4, .... -2s + 2, -2s\}$. Note that the maximal value of $\Lambda^z(s)$ is $2s$. Thus, we scale and define 
\begin{eqnarray}
C^{zz}_{1:\text{rest}} = \frac{1}{2s}|\text{tr}(\rho_{12...N}~\sigma^z_1 \otimes \Lambda^z_{23...N}(s))|.
\end{eqnarray}
Having laid out the tools, we now compute the monogamy score for $\delta_{C^{zz}} = C^{zz}_{1:\text{rest}} - \sum_{i=2}^N C^{zz}_{1i}$ for $N = 3, 4, 5$ and $6$. Our investigations from the frequency distribution of  $\delta_{C^{zz}}$ reveal that all randomly generated states are nonmonogamous irrespective of the values of $N$. Moreover, with increase of $N$,   $f$-distribution of monogamy scores also does not change much and as mentioned, all the randomly generated state remain nonmonogamous, i.e., ubiquitously follow a polygamy relation. Furthermore, note that our conjectured $\Lambda^z(s)$ cannot be written as a sum of local magnetizations $\oplus_{i = 2}^N \sigma^z_i$.
This suggests that our proposed bound, as inspired from monogamy, is not a particularly good one in this case, as also depicted in Fig. \ref{fig:mscc1}. We will contrast the results with classical  discord and local work in the subsequent subsection.

\subsection{An upper bound for CQD and LW from monogamy}

When monogamy-based upper bounds, $C_{1:\text{rest}}^{D/LW}$,  on \(\sum C_{1i}^{D/LW}\) are employed in case of the classical part of QD and local work, it seems to work much better compared to the case of classical correlators, especially when the random states contain more number of qubits.
 The analysis shows yet another point of qualitative difference between the usual classical correlators and the axiomatic classical correlation measures, see Fig. \ref{fig:mscc1}.

We track the quality of the bounds by examining the $f$-distribution of the monogamy scores and by computing its statistical parameters of the distribution, see Tables. \ref{table:disggm} and \ref{table:defggm}. In particular, we are interested in the percentage of states that satisfy the monogamy inequality, i.e., the percentage of states for which $\delta_{C^{D}}\geq 0$ and   $\delta_{C^{LW}}\geq 0$. Since both classical discord and local work behave almost identically, we list our general observations for both these quantities below:
\begin{enumerate}
\item \emph{Mean  and standard deviation of monogamy score.} Unlike classical correlators, the mean monogamy score progressively shifts from negative to positive values on increasing $N$ from $3$ to $6$ while the standard deviation does not follow any strict pattern in these cases (see Fig. \ref{fig:mscc1}).

\item \emph{Percentage of states satisfying monogamy.} For $N= 3$, we find that only a few states satisfy the monogamy relation. However, as $N$ is increased to $6$, almost all random states $(\sim 99\%)$ satisfy the monogamy relation. It suggests that our imposed monogamy-based bound works better when the number of qubits in the generated random states grows. Here it is important to note that monogamy score for QD and WD also increases with $N$ and reaches close to maximal value with the increase of $N$ \cite{Sooryarand, mono_app4}.

\item \emph{Connecting monogamy-based bound with genuine multipartite entanglement.} Furthermore, if one looks at the data from the $f$-distribution of the classical  discord and local work by laying it out on the grids of genuine multipartite entanglement content of the random pure states, we observe an interesting feature. Specifically, when $N$ increases, we know that random states that possess more genuine multipartite entanglement  on average \cite{Eisertrand, Winterrand}. We observe a strong correlation of the GGM enhancement as  $N$ increases, with proclivity of a major percentage of randomly generated states satisfying the monogamy relation for axiomatic CC measures as depicted in Fig. \ref{fig:mscc}, i.e., high genuine multipartite entangled states satisfy the monogamy of CQD and LW.  
\end{enumerate}

\begin{table}
\begin{center}
\begin{tabular}{|c|c|c|c|c|}
\hline
$N$ &  3 & 4   & 5 & 6 \\
     \hline
mean &  -0.254  & 0.0172 & 0.344 & 0.593   \\     
\hline
sd & 0.190 & 0.272 & 0.113 & 0.074 \\ 
\hline
$\mathcal{M}$ & 6.792 & 54.606 & 99.458 & 100.00  \\
\hline
\end{tabular}
\end{center}
\caption{Mean and standard deviation (sd) of the distribution for the monogamy score of the classical part of discord, \(\delta_{C^D}\) with a step size of 0.01. \(\mathcal{M}\) denotes the percentage of monogamous states obtained from randomly generated states. The total number of random states simulated for the analysis for each N is \(10^5\).}
\label{table:disggm}
\end{table} 

\begin{table}
\begin{center}
\begin{tabular}{|c|c|c|c|c|}
\hline
$N$ &  3 & 4   & 5 & 6  \\
     \hline
mean &  -0.182  & 0.042 & 0.310 & 0.522  \\     
\hline
sd & 0.145 & 0.155 & 0.121 & 0.075  \\ 
\hline
$\mathcal{M}$ & 7.154 & 65.835 & 98.264 & 99.998  \\
\hline
\end{tabular}
\end{center}
\caption{Similar analysis as in Table \ref{table:disggm} is performed for the monogamy score of LW, \(\delta_{C^{LW}}\). }
\label{table:defggm}
\end{table}

\section{Conclusion} 
\label{sec_conclu}

In a multipartite state, shareability of quantum correlations (QC) among its two-party subsystems is restricted  while such a distribution of classical correlations (CC) among parties  is not forbidden.   In particular, classical correlation content can be maximum simultaneously for all the bipartite reduced density matrices  of a multipartite state. It raises a natural question whether all states chosen Haar uniformly  from a state space also possess the similar feature. Specifically, our aim was to find out the shape of  the distribution for the sum of  CC measures obtained from  the reduced density matrices of random multipartite states. We also addressed the question whether the  maximum value for shareability of CC  is different for random states than the one obtained via a class of states or not.   

To investigate  it, we considered three kinds of classical correlation measures -- conventional classical correlators, CC measure appearing in the definition of quantum discord and extractable local work in quantum work-deficit. The last two definitions of CC measures obey certain axioms while the first one arises from the measurements performed on two spatially separated systems. Our results showed that although these axiomatic classical correlation measures have some distinct dissimilarities with classical correlators,  the overall behavior of these measures  follow a uniform  
pattern. To study the behavior, we have chosen two directions -- we considered the pattern of  the distributions obtained for the sum of a given classical correlation measure distributed among two-parties of random multipartite states and  we call the situation as unconstrained one; secondly, we studied the distribution of classical correlation measures when the states possess a fixed amount of other classical correlation or genuine multipartite entanglement, referred as the constrained scenario. For our analysis, we generated Haar uniformly random three-, four-,  five- and six-qubit states. 
 In the unconstrained case, we found that their distributions have Bell-like shape with one long-sided tail, and the mean of the distributions is almost constant for classical correlators with the increase in the number of parties while the average values of the distribution for the axiomatic CC measures decrease when the number of qubits vary. In case of classical correlators, we also showed that the noncommutativity in the directions on which classical correlators are defined played an important role in the pattern of shareability of classical correlators. 

In the constrained case, we observed that average and maximum values of shareability for conventional  classical correlators  does not depend on the genuine multipartite entanglement content  although two noncommuting classical correlators depend on each other. Interestingly, we found that for a given genuine multipartite entanglement, maximal value of local work and classical part of quantum discord showed a dual nature in a sense that when one increases, the other one decreases, especially for three-party states.  

Counter-intuitively, we observed that  the maximal value of CC  measures, both from the axiomatic  and the conventional one, of random multipartite states can be far from the algebraic maximum that CC measures can reach for a certain class of states. Such an observation tempted us to check whether the monogamy-based  bound can also be an upper bound for CC measures. 
We believe that the results obtained here reveal a distinct rule for the distributions of classical correlation measures among subsystems of a global multipartite system. These restrictions are different from the constraints in shareability known for quantum correlation measures.

\acknowledgements
We acknowledge the support from Interdisciplinary Cyber Physical Systems (ICPS) program of the Department of Science and Technology (DST), India, Grant No.: DST/ICPS/QuST/Theme- 1/2019/23. Some numerical results have been obtained using the Quantum Information and Computation library (QIClib). This research was supported in part by
the INFOSYS scholarship for senior students. We also thank the anonymous Referees for  insightful suggestions.


\begin{thebibliography}{100}
\bibitem{GGMdef} The generalized geometric measure is a distance-based computable multipartite entanglement measure for pure states which is defined as the distance between a given multipartite pure state and a closest nongenuinely multipartite entangled state. It can be computed if one finds the maximum from the set containing all  the maximum eigenvalues obtained from all possible bipartitions of a \(N\)-party state, \(|\psi_{12\ldots N}\rangle\).
\end{thebibliography}
\end{document}